\documentclass{aastex}

\begin{document}
\doublespace
\title{A New Mechanism for Chondrule Formation: \\ Radiative Heating by Hot Planetesimals }

\author{William Herbst}
\affil{Astronomy Department, Wesleyan University, Middletown, CT
06459}
\email{wherbst@wesleyan.edu}

\and

\author{James P. Greenwood}
\affil{Earth \& Environmental Sciences Department, Wesleyan University, Middletown, CT
06459}
\email{jgreenwood@wesleyan.edu}

\begin{abstract}

We propose that chondrules are formed by radiative heating of pre-existing dust clumps during close fly-bys of planetesimals with incandescent lava at their surfaces. We show that the required temperatures and cooling rates are easily achieved in this scenario and discuss how it is consistent with bulk aspects of chondritic meteorites, including complementarity and the co-mingling of FeO-poor and FeO-rich chondrules. 
  
\end{abstract}
 

Chondrules are small (0.1-2 mm) spheres of igneous rock that make up 30-80\% of the volume of primitive meteorites. It is generally agreed that they formed 1 - 3 Myr after the CAIs \citep{k12,k13}. Their textures require heating to peak temperatures of 1750-2100 K, but not above, and for only a few minutes \citep{g96,h96}. Cooling rates are inferred to be rapid, but not too rapid, typically 100-1000 K/hr \citep{hr90,h05,d12}. What could possibly provide the appropriate dose of heat at just the right time and in just the right manner? This question has been labeled as ``arguably the the major unresolved question in cosmochemistry'' \citep{phe15}.

\citet{c05} reviewed theories of a decade ago, categorizing them as {\it nebular} and {\it planetary}. Nebular theories, which probably remain most popular today, assume that pre-chondrule dust clumps are entrained in a H/He gaseous disk and heated by some process that deposits energy in the gas. A variant of this, the X-wind model, uses the gas component of the disk to transport the pre-chondrule material close enough to the Sun to reach the required temperatures and then back to $\sim$3 AU for incorporation into the chondrite meteoroid. The popularity of these theories comes in part from their ability to predict thermal histories that match fairly well what laboratory experiments tell us is required to form igneous spheres with the observed, mostly porphyritic, textures of actual chondrules. \citet{d12} compare the experimental results with predictions of leading nebular models and conclude that pressure shocks induced by gravitational instabilities within the gaseous disk do a better job than other leading nebular theories, such as planetesimal bow shocks, lightning, or the X-wind model, in matching inferred thermal histories.  

A serious challenge to all nebular theories comes from the elegant work of \citet{a08}. These authors show convincingly from the distribution of Na$_2$O in many chondrules that these objects could only have formed at very high densities of solids, $\rho_s$, in the range $\rho_s \approx 10^{-5}-10^{-3}$ gm cm$^{-3}$. Since the surface density of solids ($\Sigma_s$) in the minimum mass solar nebula is $\Sigma_s \approx 10$ gm cm$^{-2}$, to achieve those number densities requires compressing the disk to a full-width thickness (W) of $W = \Sigma_s \slash \rho < 10$ km. Obviously this is difficult if the solids are still embedded in a gaseous nebula at the time they are heated to chondrule formation temperatures. In addition, \citet{f12} show that significant (unobserved) isotopic signatures would be expected in chondrules formed under any plausible nebular conditions. Suppressing these signatures requires higher solids density, higher gas pressure and/or shorter heating times than nebular shock models can produce. 

Another challenge comes from the increasingly accurate measurements of chondrule and CAI ages by \citet{c12} and \citet{b14}. These authors confirm that the epoch of chondrule formation is significantly longer than the epoch of CAI formation, continuing for at least 2-3 Myr. They further substantiate that there was a delay in the formation time of most chondrules of about 1 Myr from t = 0, as measured by the CAIs. This demonstrates that the heating event associated with CAI formation, normally taken to be the hot nebular phase expected as infalling gas from the protostellar envelope dissipates its kinetic energy in transitioning to an accretion disk, is not the same heating event (or events) responsible for chondrule formation. Delaying the main chondrule heating event by 1 Myr and extending it for another 2 Myr before shutting it down forever is a serious challenge to all hypotheses, nebular and planetary alike. 

Planetary models invoke collisions between planetesimals to form chondrules. The complexity of such a process has stymied our ability to prove by numerical simulation or other means that chondrules with the observed range of properties could actually be made in this manner. While gaining some popularity recently, perhaps because of the difficulties being encountered by nebular models, planetary models have had relatively few adherents. Two obstacles to any successful planetary model, raised by \citet{c05}, bear repeating: 1) collisions occur throughout the history of the solar system, yet the chondrule formation time is clearly demarcated, and 2) there is evidence for multiple heating of chondrules. 

A further challenge to planetary models comes from observations of the phenomenon known as {\it complementarity}. Recent studies \citep{e08, phe15} extend the evidence along lines first presented by \citet{w63}. It is found that, while bulk compositions of carbonaceous chondrites are consistently solar, chondrules have low Si/Mg and Fe/Mg ratios, while matrix has high ratios. The authors argue that this reflects pre-accretionary processes and conclude, therefore, that the chondrules and the matrix must have formed from a single reservoir of solar composition. An independent origin of chondrules followed by transport and mixing with matrix from elsewhere prior to formation of the meteorite parent body is excluded. At face value, this rules out the X-wind jet nebular hypothesis and severely complicates any planetary model. Nonetheless, recent versions of both nebular and planetary models have been developed, some with the intent of addressing these challenges. On the planetary side there is the idea of collisions of already molten planetesimals proposed by \citet{ss12} and the impact jetting model of \citet{j15}. On the nebular side, there are modern versions of the pressure shock hypothesis \citep{hw12} and current sheets \citep{m13}, but none successfully meets all the challenges of the data.  

Here we propose a new mechanism for chondrule formation that is neither nebular nor planetary. We assume, as in nebular theories that chondrules are formed from pre-existing dust clumps held together by electrostatic force that are probably part of larger aggregates. But we take seriously the evidence that, by the time the chondrules are heated - producing igneous spheres of the observed textures - there is little or no nebular gas remaining in their environment. Whatever gas is present comes from the evaporation of volatiles within what we infer to be the pre-chondrite aggregate. The heating agent is radiation from hot lava at the surfaces of planetesimals, probably in the 10-100 km size range. 

It is known from the properties and ages of iron meteorites that relatively small (r $\sim$50 km) molten planetesimals were present in the forming solar system within the first $\sim$1 Myr of its existence \citep{h90,g05,s06}. The most likely heat source for these objects is $^{26}$Al. As \citet{l76,lw77} first showed, its inferred abundance at t = 0 is sufficient to melt planetesimals larger than a few km on a time scale of $\sim$3 $\times 10^5$ yr. More recent and detailed models by \citet{ss12} confirm this scenario, finding that planetesimals with r$\ge$10 km will fully melt within 0.3-2.5 Myr. Based on these studies we may reasonably conclude that there was an abundance of magma present in the pre-planetary disk at the time of chondrule formation, encased in 10-100 km radii planetesimals, where it had been incubating since the epoch of planetesimal formation. From time to time some of this molten material must have reached the surfaces of these significantly overheated planetesimals, resulting in crustal foundering and the appearance of incandescent lava at the surface. Even today in the Solar System we regularly see hot spots at the surface of the most overheated body, Io \citep{k07}. Could the radiation from hot surface lava and exposed magma oceans \citep{g05} -- plausibly be the heat source for melting pre-chondrule dust aggregates within highly solids-enriched aggregates? Fortunately, this proposal has sufficiently simple physics that we can robustly predict a thermal history for the chondrules if this is how they formed. As we now show, that thermal history is both independent of the size of the planetesimal providing the heat and an intriguingly good match with what the textural data require.

To quantitatively assess the plausibility of this idea we consider a fully molten planetesimal of density, $\rho$, radius, r, and surface temperature, T$_s$, where to be useful for chondrule heating, T$_s$ will need to be equal to, close to, or above the maximum liquidus temperature for chondritic material. For the purposes of this exercise, we will adopt  $\rho$ = 3 gm cm$^{-3}$ and T$_s$ = 2150 K. The choice of density affects the shape of the cooling curves while the choice of T$_s$, which is based on calculations reported by \citet{hr90}, affects how close a fly-by of the surface is needed to heat material sufficiently for chondrule formation. We neglect reflective losses, assuming that all of the radiant energy incident on the dust aggregate will be absorbed within it. Our initial focus is on the individual dust clumps that will form chondrules. Their aggregation, which may form the chondrite, could influence the heating/cooling relation by lengthening the time it takes to cool, but discussion of that point is beyond the scope of this exploratory study. 

Material passing close enough to the surface of an object with an exposed magma ocean can get heated by radiation to the sub-liquidus temperatures required for chondrule formation. Obviously, the material cannot be heated above the liquidus temperature and since the dust would need to be {\it very} close to the surface to reach the liquidus, it is clear that in a scenario like this, a sub-liquidus temperature will be the norm -- as required of any chondrule formation theory. Chondrule-mass objects will quickly come into equilibrium with the radiation field and we assume they do so instantaneously. To estimate cooling rates, we consider the case of a parabolic orbit and neglect gas drag and any effects of re-radiation from surrounding chondrules. Small planetesimals in the 10 -100 km range are unable to hold an atmosphere, although perhaps outgassing of volatiles from the magma or evaporation of the silicates could produce an expanding ``exosphere'' that should be factored into any more detailed model in the future. Hydrogen and helium associated with the solar nebula may be safely ignored at the high density of solids scenario envisioned here. In this exploratory study we, therefore, need consider only the dynamical effects of gravity. 

\begin{figure}
\epsscale{1.0}
\plotone{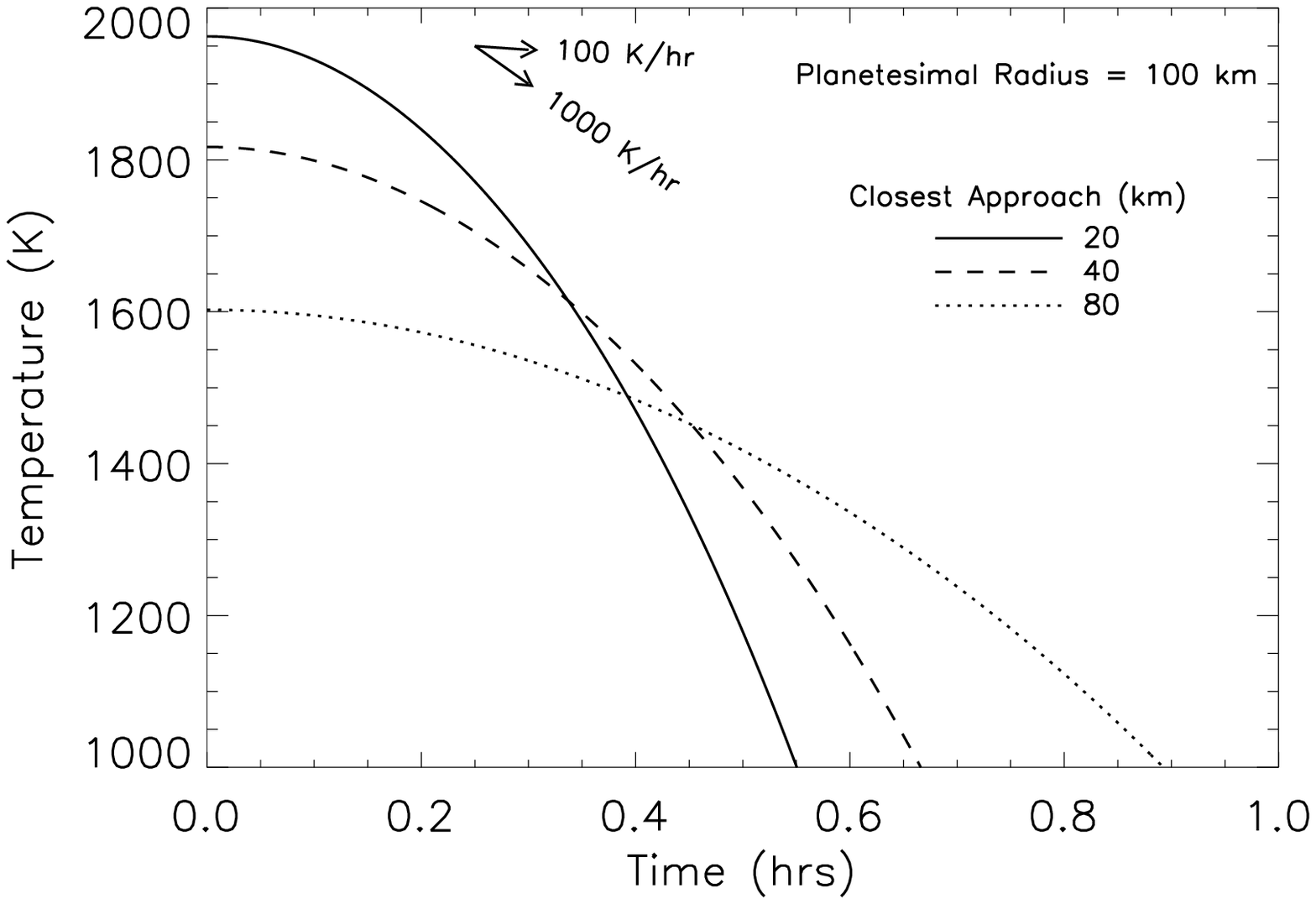}
\caption{Cooling curves for chondrules from our model compared to observed rates of 100-1000 K/hr. These rates are independent of the planetesimal radius, depending only on its density (see Fig. \ref{fig4}).}
\label{fig3}
\end{figure}

\begin{figure}
\epsscale{1.0}
\plotone{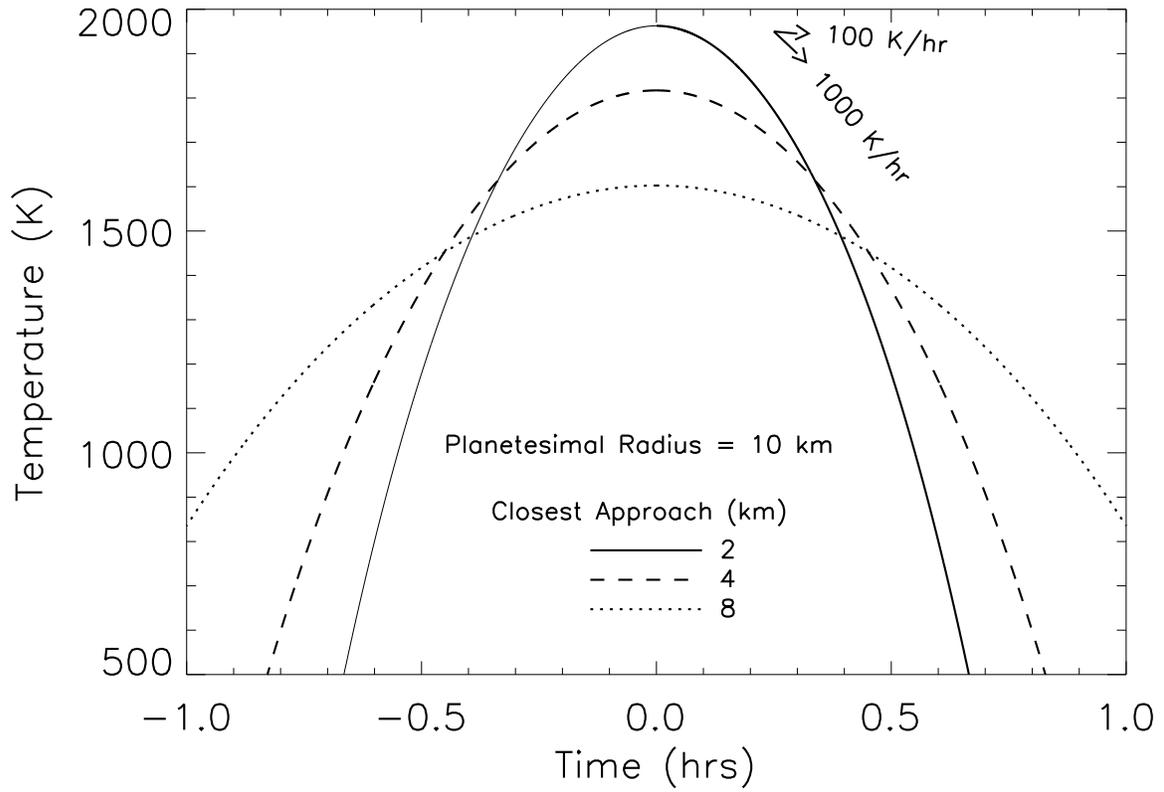}
\caption{Same as Fig. \ref{fig3} but showing a larger range of time, including the pre-heating phase. These calculations were done for a 10 km radius planetesimal but, again, the form of the curves is independent of the size of the planetesimal.}
\label{fig4}
\end{figure}

The resulting heating and cooling curves are shown in Figs. \ref{fig3} and \ref{fig4}. It is evident that, with the simplest of assumptions and underlying physics, we obtain a predicted thermal history for this model that can be directly compared with observational constraints. As \citet{d12} discuss, constraints on the thermal history of chondrules are the most powerful tool for assessing viability. An important aspect of our model is that porphyritic chondrules, the most abundant texture, require subliquidus temperatures to form \citep{l96} and that is precisely the temperature range the model predicts. The examples shown are for a planetesimal of radius 100 km (Fig. \ref{fig3}) and 10 km (Fig. \ref{fig4}) but the peak temperature and shape of the curves depends only on the ratio x = h/r. Smaller planetesimals provide less acceleration to the passing dust but they also heat smaller volumes and these effects offset each other to produce heating and cooling curves that depend only on the ratio, x, not on r. It is clear from these figures that a simple model of radiative heating by close passage to a planetesimal with a molten lava surface, under the action of gravity alone, accounts for both the peak temperatures and basic cooling rates required to explain the formation of chondrules. 

There are two ways in which the model thermal history presented in Fig. \ref{fig4} differs from the most successful nebular models, pressure shocks, discussed by \citet{d12}. First, there is no ``flash heating" but, instead, a symmetric (with the cooling) and more gentle rise in the temperature as the pre-chondrule material approaches the hot lava surface. Second, the simple model explored here has a generally more rapid cooling rate at later times than the pressure shock models, with temperatures remaining above 1400 K for only an hour or so, compared to one or two days in the pressure shock scenarios. These differences could be significant in testing our hypothesis in the future; compared to other models there does not seem to be much latitude in the thermal history we can accommodate, an aspect of our model that we consider to be a strength. If chondrules are proven to require a heating history significantly different from what is shown in Fig. \ref{fig4} to form, then our model may be ruled out. At present, the issue is debatable because the main argument for flash heating is the presence of volatiles within chondrules but if volatiles can evaporate during heating and re-condense during cooling, then the constraint of flash heating is relaxed \citep{a08,d12}. Slower cooling rates at later times could be generated in our scenario by including radiation from surrounding matter, but it is not entirely clear to us that experimental data currently requires cooling rates significantly below $\sim$200 K/hr. We believe, therefore, that the predicted thermal history for this model is consistent with all current constraints on chondrule formation and that future developments on the experimental side may eventually strengthen or falsify our model. 

The heating model proposed here also provides a natural explanation for processes requiring a range of temperatures seemingly within the same small volume of space, such as the formation of two chemical types of chondrules -- FeO-poor and FeO-rich. Pre-meteorite material further from the surface magma will be cooler than that closer to it for two reasons: 1) the proximity effect, and 2) attenuation of the radiation field by the dust closer in. If the ambient radiation field responsible for heating material is attenuated by a factor $e^{-\tau}$ then the equilibrium temperature reached by the solids will be reduced by a factor $e^{-\tau / 4}$. At an optical depth of $\tau = 1$ the equilibrium temperature is reduced by 22\%, corresponding to several hundred degrees. This is sufficient to account for the difference between chondrule types. The path length within the aggregated dust that is required to cause such attenuation depends on the opacity of the material. We can estimate the path length corresponding to $\tau = 1$ as follows. Assume that the matter is concentrated into spheres of radius, r, and grain density, $\rho_g$. The optical depth is given by $\tau = k l$, where k is the volumetric opacity; k$^{-1}$ is the mean free path, $l$, of a photon in this material. For absorbers of a single size we may write that $k = n \sigma$ where n is their number density and $\sigma$ is the extinction cross-section per absorber. Following Mie theory, we take $\sigma = 2 \pi r^2$. Calculating n as $\rho_s / m$ where m is the mass of each dust grain yields
$$ {l = k^{-1} = {{2 r \rho_g}  \over 3 \rho_s}} $$
It seems safe to say that $r \le 0.1$ cm and $\rho_g \le 3$ gm cm$^{-3}$, so $l \le 0.2 / \rho_s$. At a typical density for chondrule formation based on the Na$_2$O results of \citet{a08}, namely $\rho_s \approx 10^{-4}$ gm cm$^{-3}$, $l \le 20$ meters. If chondrules typically require such densities of solids and if the radiative heating mechanism proposed here operates, one will expect to have a wide temperature range within the formation zone. Of course when full aggregates are considered instead of individual dust clumps, there will be some moderation of these effects.

The discussion above suggests that the size of the chondrule formation zone, L, may be much smaller than has previously been considered in nebular models. For example, \citet{d12} argue that L $\gg$ 10$^3$ km is required, based on the percentage of chondrules (5\%) that are compound, indicating that they collided while still hot. Adopting the same numbers as in the previous paragraph, and r = 0.03 cm, we can calculate the number density of chondrules in the formation zone as n = ${\rho_s \over 2m} \approx 0.12$ cm$^{-1}$, where we have assumed that half of the solids mass is in matrix material. This, in turn, implies a mean free path for the chondrules of $L = (n \pi r^2)^{-1} = 30$ m. The constraint that 5$\%$ of chondrules suffer a collision while hot would be satisfied if their relative velocity, v$_{rel}$, satisfied the constraint $n \pi r^2 v_{rel} t = 0.05$, where t is the duration of the chondrule hot phase. Based on Fig. \ref{fig4} we take t = 1 hr, which yields $v_{rel} = 0.04$ cm s$^{-1}$. This velocity is sufficiently low that there seems little doubt that it could be achieved. Clearly, our picture of chondrule formation suggests a more intimate link with the process of chondrite formation than is the case for nebular models.

The expected temperature range over the chondrule formation zone can account for the fact that FeO-poor (Type I) chondrules and FeO-rich (Type II) chondrules are found intermingled in chondritic meteorites. The FeO-rich chondrules are more oxidized, have more volatiles, have lower liquidus temperatures and have higher $\Delta^{17}$O compositions than the FeO-poor chondrules \citep{a08,u12,s13}. It has been challenging for models of chondrule formation to account for these fundamental chemical differences in the two main types of chondrules \citep{v12}. That these two types of chondrules, which apparently formed in different environments, are found intimately mixed in the chondritic meteorites, has required separate formation locations followed by physical mixing in some later process such as \citet{ss12}, for example, have proposed. On the other hand, the constraint of complementarity requires that the FeO-rich and FeO-poor chondrules, and their associated matrix, formed in close proximity to one another \citep{phe15}. We note further that volatiles lost from the most intensely heated (inner region) of the material could have led to an enhancement of the local oxygen fugacity and other volatiles, such as Na, within the cooler, outer zone. Minor incorporation of oxygen from nebular ices into the FeO-rich chondrules can account for their oxygen isotope differences from FeO-poor chondrules, as nebular ice is expected to be highly enriched in $\Delta^{17}$O \citep{s07}. 

Can our proposed mechanism account for the large volumetric abundance of chondrules within chondritic meteorites? Yes, if we presume that chondritic meteorites form during the same heating events that form the chondrules. Gravitational focusing will lead to a significant compression of the incoming material. Coupled with the heating experienced by the chondrules and matrix one has the ingredients needed for bulk meteorite formation. Fine grained dust, such as that which composes the matrix, might survive radiative heating at 2000 K since it is too small to efficiently absorb energy emitted near the peak of the black body curve. Coarser grains will evaporate under conditions where highly refractory material and very fine grained material might survive, possibly accounting for the basic component structure of chondrite meteorites. If the heating and compression required for meteorite formation involves a sufficiently close passage past a hot planetesimal that chondrule formation inevitably occurs, then the fact that chondrules are so abundant in primitive meteorites is explained. We note that, if this line of argument is correct, it means that chondrules are an abundant component of primitive meteorites, but not of the early solar system in general.
  
To summarize, we propose that chondrules are formed when pre-existing dust clumps, probably within aggregates, are heated during close encounters with incandescent lava at the surfaces of planetesimals and that chondritic meteorites are probably often formed in the same heating and compression events. The model accounts or potentially accounts for all of the well-established features of chondrules and chondritic meteorites, notably including their thermal history, age range, complementarity, formation environment and ubiquity within chondrites. In particular, we note the following:

1. Experimental data indicates that chondrules formed from material heated to peak temperatures of at least 1750 - 1820 K, but not exceeding 2100 - 2370 K, for times of order minutes, which then cooled at rates of hundreds or, at most, a few thousands of degrees per hour. As Figs. \ref{fig3} and \ref{fig4} show, this is precisely the thermal history our model predicts for material passing within about 0.5 planetesimal radii of the surface, regardless of the size of the planetesimal involved. Unlike nebular models, we do not predict a highly asymmetric ``flash" heating. In our model, the retention of volatiles and absence of large isotopic anomalies is due to the high vapor pressures associated with a high density of solids in the pre-chondrite aggregate and rather brief, overall, heating times.   

2. The epoch of chondrule formation is readily understood in our model as the time in the history of the Solar System when molten lava was  present in abundance throughout the terrestrial planet forming zone. The peak of this epoch is $\sim$1 Myr after CAI formation, because it typically takes that long to incubate the energy available in $^{26}$Al to temperatures $\ge$ 2000 K within the interiors of planetesimals with radii of $\sim$10 km or larger. The epoch is over by $\sim$3 Myr because the decay of the $^{26}$Al fuel relieves the condition of overheating; molten lava becomes increasingly less common at the surfaces of small planetesimals thereafter.

3. Multiple heating events for the same chondrules, as observed, are certainly possible. Not every dust clump or aggregate of them that is heated during close passage to a hot surface needs to form a chondritic meteorite, although most will if complementarity is a widespread feature of chondrites.   

4. The model potentially meets the constraints of complementarity since it has a local heat source acting on a reservoir of bulk solar composition. Newly formed chondrules and heated matrix from the same original reservoir may be concentrated within the gravitational wake of hot planetesimals, which we propose as a possible site of chondritic meteorite formation.

5. The radiative heating model easily accommodates the observation that chondrules are formed at a density of solids, $\rho_s \approx 10^{-5}-10^{-3}$ gm cm$^{-3}$, since no nebular gas whatsoever is required in the heating process.  The existence of FeO-poor and FeO-rich chondrules within the same meteorite may be understood as a consequence of the expected temperature range in the pre-meteorite aggregate caused by variable proximity to the surface radiation and the relatively high opacity in the near-infrared. The high opacity of the material limits the chondrule formation zone to tens of meters, consistent with the frequency of compound chondrules. 

6. Chondrule formation experiments should be useful in testing the rather highly constrained and symmetric heating and cooling curves predicted by our model, which are different in significant ways from the predictions of nebular models. 

\acknowledgments

We thank Denton Ebel and his staff at the American Museum of Natural History for their hospitality during visits by each of us and for many helpful lessons on chondrules and chondrites. We thank the editor and referees of this paper for helpful comments during the review process.

\clearpage

\end{document}